\documentclass[aps,pra,twocolumn,showpacs,
floatfix,nofootinbib,groupedaddress,superscriptaddress,citesort]{revtex4}
\usepackage{mathrsfs}
\usepackage{amsfonts}
\usepackage{amstext}
\usepackage{amsmath}
\usepackage{amssymb}
\usepackage[dvips]{graphicx}

\begin{document}

\title[Short title for running header]{Spin-orbit coupling induced half-Skyrmion excitations in rotating and rapidly quenched spin-1
Bose-Einstein condensates }

\author{Chao-Fei Liu}
\affiliation{School of Science, Jiangxi University of Science and Technology, Ganzhou 341000, China}
\affiliation{Beijing National Laboratory for Condensed Matter Physics, Institute of
Physics, Chinese Academy of Sciences, Beijing 100190, China}

\author{W. M. Liu}
\affiliation{Beijing National Laboratory for Condensed Matter Physics, Institute of
Physics, Chinese Academy of Sciences, Beijing 100190, China}


\begin{abstract}
We investigate the half-Skyrmion excitations induced by spin-orbit coupling in the rotating and rapidly quenched spin-1 Bose-Einstein condensates. We give three expressions of the corresponding spin vectors to describe the half-Skyrmion. Our results show that the half-Skyrmion excitation depends on the combination of spin-orbit coupling and rotation, and it originates
from a dipole structure of spin which is always embedded in three vortices constructed by each condensate component respectively.
When both the strength of spin-orbit coupling and rotation frequency are larger than some critical values, the half-Skyrmions encircle the center with one or several circles to form a radial lattice, which occurs even in the strong ferromagnetic/antiferromagnetic condensates.
We can use both the spin-orbit coupling and the rotation to adjust the radial lattice.
The realization and the detection of the half-Skyrmions are compatible with current experimental technology.
\end{abstract}

\maketitle

\section{\textbf{Introduction}}

Spin-orbit coupling (SOC) describes the interaction of the spin of a particle with its
motion. The particular form of SOC can be of either Rashba \cite{Bychkov} or
Dresselhaus \cite{Dresselhaus} type. SOC in an electronic system \cite{ Niuqian} is able to serve as a spin
filter or a Stern-Gerlach apparatus. And it is crucial
for the spin-Hall effect \cite{Kato, Konig} and topological
insulators \cite{Kane, Bernevig, Hsieh, LiuVincent}.
Recently, spin-orbit coupled Bose-Einstein condensate (BEC) has been realized experimentally in NIST \cite{Lin}.
Unlike the previous experiment, their work has factually explored the
bosons system in the non-Abelian gauge field \cite{ Liao, Ruseckas, Osterloh, Satija, Zhu, Liu, Stanescu, Juzeliunas}.
This opens up a new avenue in cold atom physics and attracts much attention.
\par

Motivated by the experiment in NIST \cite{Lin},
several recent investigations about bosons with SOC have presented some nontrivial new structures
such as stripe phase \cite{Wang, TinLun, Jian, Sinha} and half-quantum vortex state \cite{Jian, Sinha, Huhui, wucj}.
Specially,
the combination effect of SOC and rotation on pseudo spin-$\frac{1}{2}$ BEC has been shown to be able to generate various vortex structures \cite{ xiaoqiang, Zhou}.
These impressive results enrich the phase diagram of the BEC system. However, as a new
effect on spinor BEC, it is not clear whether SOC can produce previously unknown types of
topological excitations such as new Skyrmion.
The study of this topic, on the one hand, can fertilize the novel quantum phases
in BEC system, on the other hand, can provide systematical description of
BEC system with various adjusting parameters related to SOC, rotation and quench \emph{etc}.
This provides a realizable experimental platform for various quantum
phenomena.

In this paper, we explore how SOC induces the half-Skyrmion excitation whose topological charge is $|Q|=0.5$ in the rotating spin-1 BEC.
In real experiments, the zero temperature cannot be fully achieved. So the ultracold Bose gases are only partially condensed, with the noncondensed thermal cloud providing a source of dissipation and leading to damping excitations. Meanwhile, evaporative cooling is a critical operation to obtain the condensate. Thus, it is necessary to refer to the finite temperature effect and a quenching process.
Furthermore, the rotation is a good tool to examine the excitations in spinor BECs \cite{ Schweikhard, aMizu, Martikainen}.
Here, we will just consider the quenching and the rotation process
on the spin-1 BEC with SOC.
We find that the half-Skyrmion excitation (non-meron-antimeron pair) is related to a three-vortex structure caused by SOC.
We give a representation of the half-Skyrmion.
The phase diagram of this system is plotted in the plane of rotation frequency and SOC strength.
We find that the generation of the half-Skyrmion excitation
must depend on the combination of SOC and rotation.
Without SOC, the Skyrmion excitation with topological charge $|Q|=1$ occurs in the rotating ferromagnetic BECs.
When both the strength of SOC and rotation frequency are larger than some critical values, the half-Skyrmions encircle the center of the system with one or several circles to form a radial lattice.
These phenomena can occur even in the strong ferromagnetic/antiferromagnetic BECs.
We can adjust the half-Skyrmions lattice by changing the strength of SOC as well as rotation.

The paper is organized as follows: In Sec. II we introduce the stochastic projected Gross-Pitaevskii equations and some initial
condition for our simulations. Sec. III is our main results
and some explanations. In Sec. III A we use the rotating spin-1 BECs
with SOC to obtain a single half-Skyrmion. We not only illuminate the essence of the
half-Skyrmion excitation, but also indicate the expressions to describe it. In Sec. III B the half-Skyrmion
lattice is studied. We further illuminate the relationship between the vortex structure and the half-Skyrmion.
We also show that the density distribution and the spin texture are dynamically stable when the system reaches the equilibrium state.
In Sec. III C the effect of the rotation frequency is discussed. Then,
we present the phase diagrams for generating the half-Skyrmions and others.
In Sec. III D we show that the half-Skyrmion can occur in the strong ferromagnetic BECs and the strong antiferromagnetic BECs with both SOC and rotation.
A summary of the paper is presented in Sec. IV.

\section{Model and equation}

Considered a quenching process in a finte-temperature BECs, the dynamics of a spin-1 BEC can
be described by the stochastic projected Gross-Pitaevskii equation (SPGPE) \cite{ Bradley, Rooney, Su}.
In the SPGPE equations, the system is divided into the coherent region and the incoherent region. The coherent region consists of all states
whose energies are below $E_{R}$ and the incoherent region contains the remaining high energy modes. This division is made by the projection operation $\mathcal{P}$.
The equations of motion are known as the form \cite{ Bradley, Rooney, Su}:
\begin{equation}
d\Psi_{j}=\mathcal{P}\{-\frac{i}{\hbar}\widehat{H}_{j}\Psi_{j}dt+\frac{\gamma_{j}}{k_{B}T}(\mu-\widehat{H}_{j})\Psi_{j}dt+dW_{j}\},
\end{equation}
where $T$ is the final temperature, $k_{B}$ is the Boltzmann constant, $\mu$ is the chemical potential, $\gamma_{j}$
is the growth rate for the $j$th component, and $dW_{j}$ is the complex Gaussian noise,
which satisfies the fluctuation-dissipation relation $\langle dW_{i}^{*}(\textbf{x},t)dW_{j}(\textbf{x}',t)\rangle=2\gamma_{j}\delta_{ij}\delta_{C}(\textbf{x}-\textbf{x}')dt$, where $\delta_{C}$ is the Dirac $\delta$ function for the condensate band field.
The projection operator $\mathcal{P}$ is used to restrict the dynamics of the spinor BEC in the coherent region.
Meanwhile, $\Psi_{j}(j=0,\pm1)$ denotes the macroscopic wave
function of the atoms condensed in the spin state $|F=1, m_{F}=j\rangle$, and
\begin{eqnarray}\label{GP2}
\widehat{H}_{j}\Psi_{j}&=&[-\frac{\hbar^{2}\nabla^{2}}{2m}+V(r)+g_{n}|\mathbf{\Psi}|^{2}]\Psi_{j} \notag \\
&+&g_{s}\sum_{\alpha=x,y,z}\sum_{n,k,l=0\pm1}(\widehat{F}_{\alpha})_{jn}(\widehat{F}_{\alpha})_{kl}\Psi_{n}\Psi^{*}_{k}\Psi_{l} \notag \\
&-&\Omega\widehat{L}_{z}\Psi_{j} +\sum_{\alpha=x,y}\sum_{n=0\pm1}\kappa_{\alpha}(\widehat{F}_{\alpha})_{jn}p_{\alpha}\Psi_{n},
\end{eqnarray}
with the coupling constants
$g_{n}=\frac{4\pi\hbar^{2}(2a_{2}+a_{0})}{3m}$, $g_{s}=\frac{4\pi\hbar^{2}(a_{2}-a_{0})}{3m}$ and the trap potential $V(r)=m\omega^{2}(x^{2}+y^{2})/2$.
$\widehat{F}_{\alpha=x,y,z}$ are the spin-$1$ matrices, $\Omega$ is the rotation
frequency, $\widehat{L}_{z}$ [$\widehat{L}_{z}=-i\hbar(x\partial_{y}-y\partial_{x})$] is the $z$ component of the
orbital angular momentum, $p$ ($p_{\alpha}=-i\hbar\frac{\partial}{\partial \alpha}$, $\alpha=x, y$) is the momentum operator, and $\kappa_{\alpha}$ denotes the strength of SOC which carries the unit of velocity.

In numerical simulations, the initial state of each $\Psi_{j}$ is generated by sampling the grand canonical
ensemble for a free ideal Bose gas with the temperature $T_{0}$ and the chemical potential $\mu_{j,0}$.
Meanwhile, the condensate band must lie below the energy cutoff $E_{R}>E_{k}=\frac{\hbar^{2}|k|^{2}}{2m}$. Noting, $k=2\pi(n_{x},n_{y})/L$, where $n_{x}$, $n_{y}$ are integers and $L$ is the size of the computation domain. To simulate the quenching process, the final temperature and the chemical potential
of the noncondensate band are altered to the new values $T<T_{0}$ and $\mu>\mu_{j,0}(j=0,\pm1)$.
Furthermore, we use the oscillator unit in the numerical computations.
The length, time, energy and strength of SOC are scaled in units of $\sqrt{\frac{\hbar}{m\omega}}$, $\omega^{-1}$, $\hbar\omega$, and $\sqrt{\hbar\omega/m}$, respectively.
In all the simulations, the trapped frequency is $\omega=200\times2\pi$, the total
number of the modes are $n_{x}$, $n_{y}=300$, the energy cutoff is
chosen at $n_{xc}$, $n_{yc}=150$, the initial temperature $T_0$ is $52nK$, the final temperature $T$ is $10nK$,
and we use $\frac{\gamma_{j}}{k_{B}T}=0.03$.

\begin{figure}[tbp]
\includegraphics[width=8.5cm]{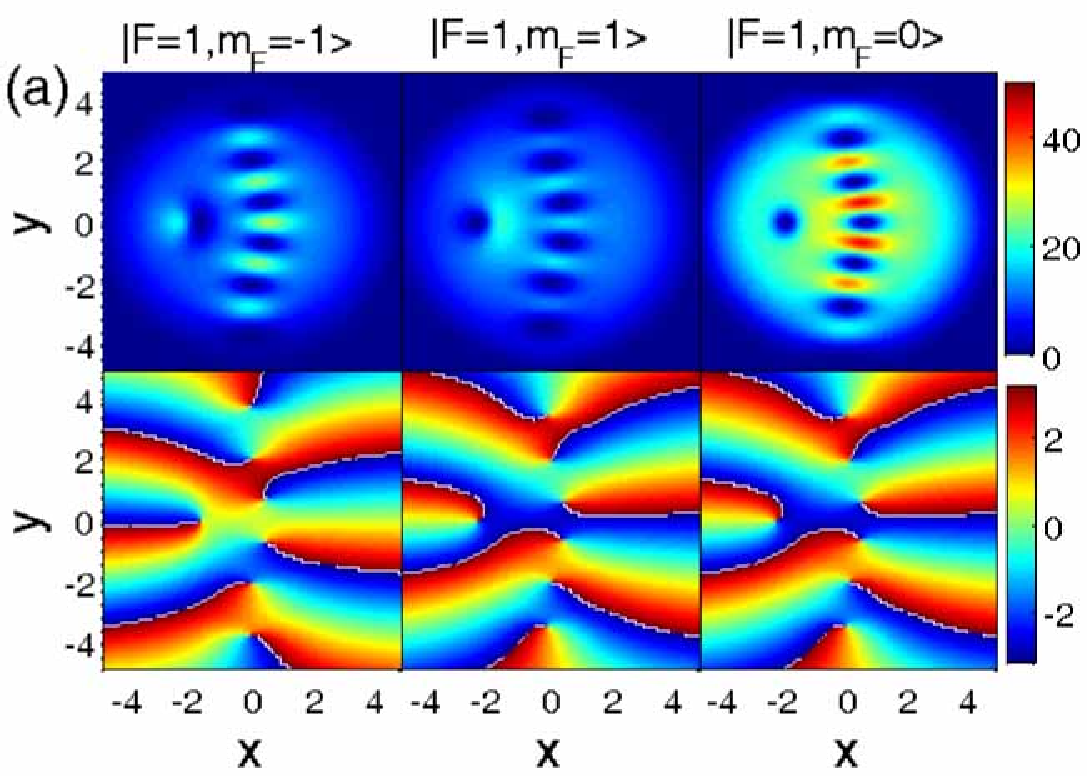}
\includegraphics[width=8.5cm]{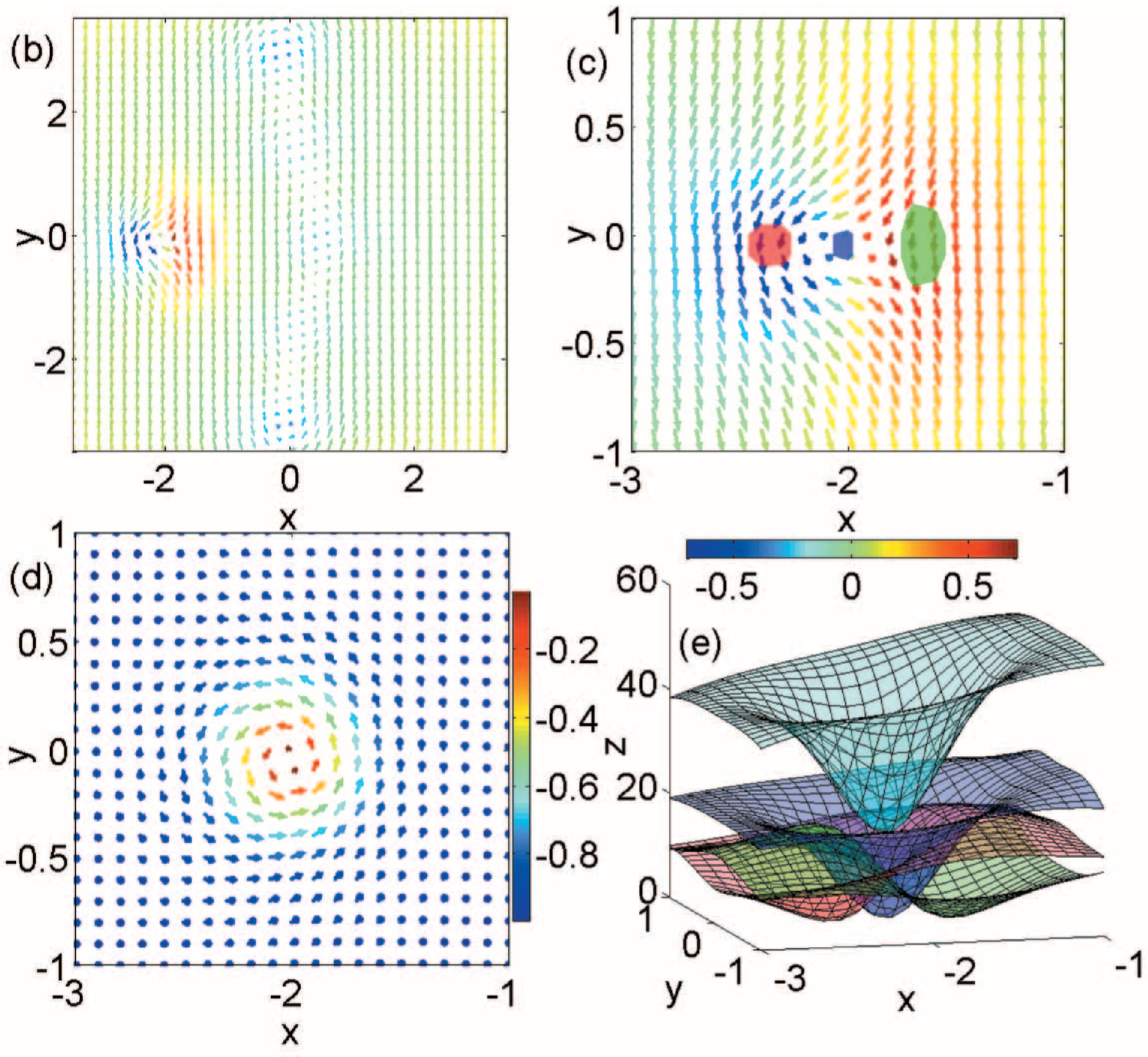}
\caption{\small (Color online). (a) The densities and phases for the spinor BEC of $^{87}$Rb when the system reaches the equilibrium state, where $\kappa_{x}=0.2$, $\kappa_{y}=1$, $\mu=8\hbar\omega$ and $\Omega=0.5\omega$. The atom numbers are $N_{-1}\approx510$, $N_{1}\approx540$ and $N_{0}\approx1050$. (b) Spin texture of the spinor BEC. The color of each arrow indicates the magnitude of $S_{z}$.
(c) The position of vortices and the spin texture.
The green,
blue and red spots are the center of vortices formed by
the $m_{F}=-1$, $m_{F}=0$ and $m_{F}=+1$ components,
respectively.
(d) Spin texture under the transformation: $(\textbf{S}^{'}_{x}, \textbf{S}^{'}_{y}, \textbf{S}^{'}_{z})=(\textbf{S}_{x}, \textbf{S}_{z}, \textbf{S}_{y})$.
(e) A scheme of three-vortex structure. The green,
blue and red surfaces denote the densities of
the $m_{F}=-1$, $m_{F}=0$ and $m_{F}=+1$ components, respectively. The cyan
is the total density of the BECs. The units of length and strength of SOC are $\sqrt{\frac{\hbar}{m\omega}}$, $\sqrt{\hbar\omega/m}$, respectively.
}
\end{figure}

\begin{figure}[htbp]
\includegraphics[width=9cm]{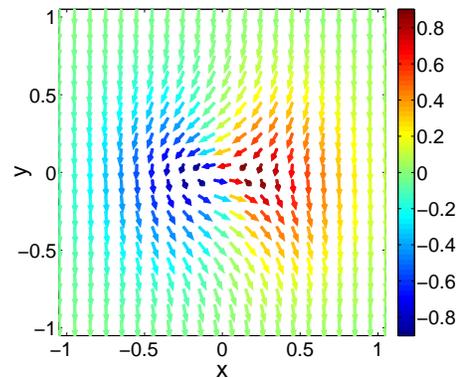}
\caption{\small (Color online). (a) Spin texture with Eqs. (4), where $\lambda=0.5$.
The color of each arrow indicates the magnitude of $\textbf{s}_{0z}$.
}
\end{figure}

\begin{figure*}[tbp]
\includegraphics[width=15.0cm]{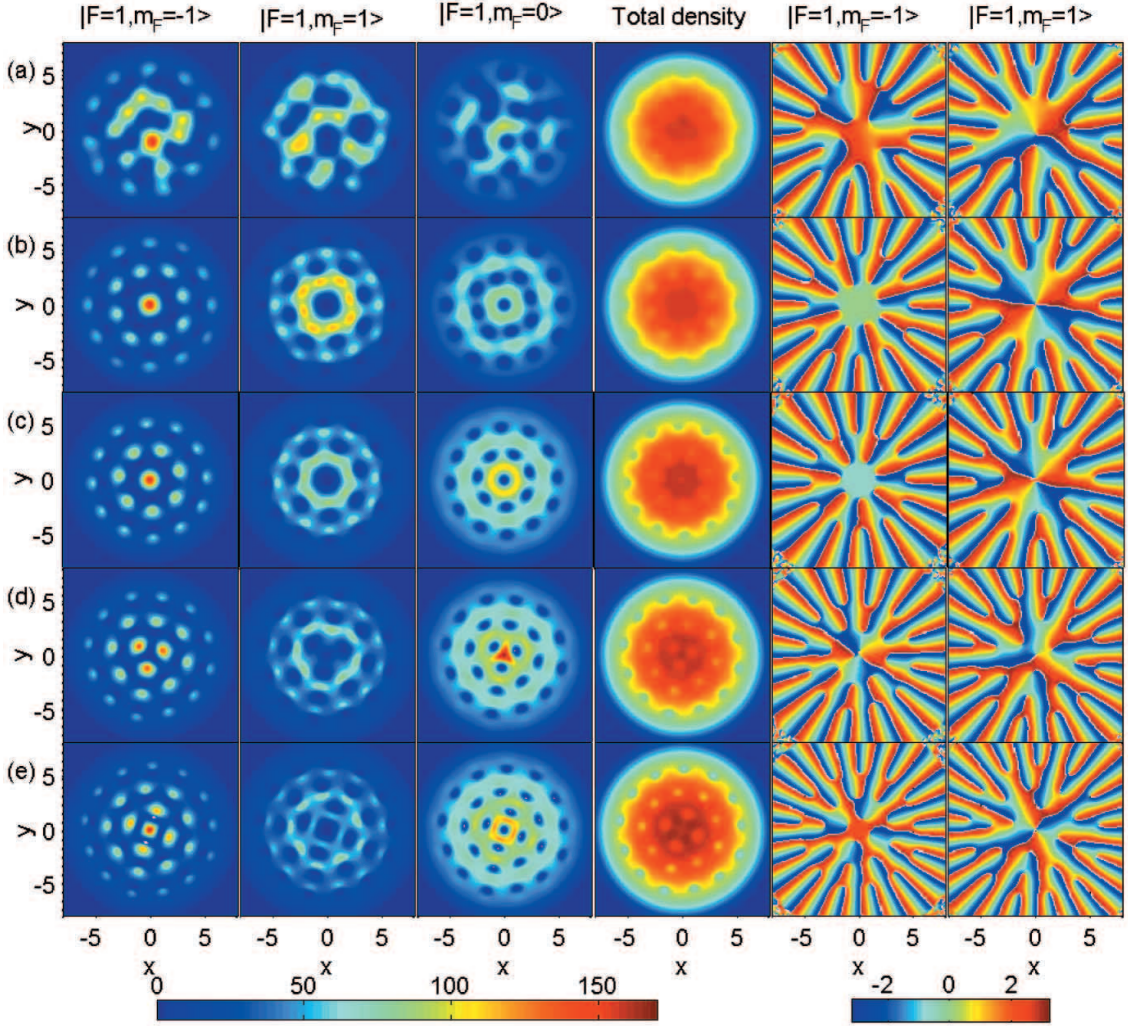}
\caption{\small (Color online). The densities and phases of the spinor BEC of $^{87}$Rb with SOC $\kappa$ when the system reaches the equilibrium state. (a) $\kappa=0.1$; (b) $\kappa=0.2$; (c) $\kappa=0.5$; (d) $\kappa=0.7$; (e) $\kappa=1.0$. Here, $\Omega=0.5\omega$, $a_{0}=101.8a_{B}$ and $a_{2}=100.4a_{B}$. Noting, the fifth and sixth columns are the phases of $m_{F}=-1$ and $m_{F}=1$ components, respectively.
The atom numbers ($N_{-1}$, $N_{1}$, $N_{0}$) approximately are ($5.1\times10^{3}$, $5.5\times10^{3}$, $5.7\times10^{3}$), ($4.0\times10^{3}$, $5.8\times10^{3}$, $6.8\times10^{3}$), ($4.0\times10^{3}$, $5.5\times10^{3}$, $8.1\times10^{3}$), ($4.4\times10^{3}$, $5.4\times10^{3}$, $8.9\times10^{3}$), and ($4.8\times10^{3}$, $5.6\times10^{3}$, $9.8\times10^{3}$), respectively. The units of length and strength of SOC are
$\sqrt{\hbar /(m \omega)}$, $\sqrt{\hbar\omega/m}$, respectively.}
\end{figure*}

\begin{figure*}[tbp]
\includegraphics[width=14.5cm]{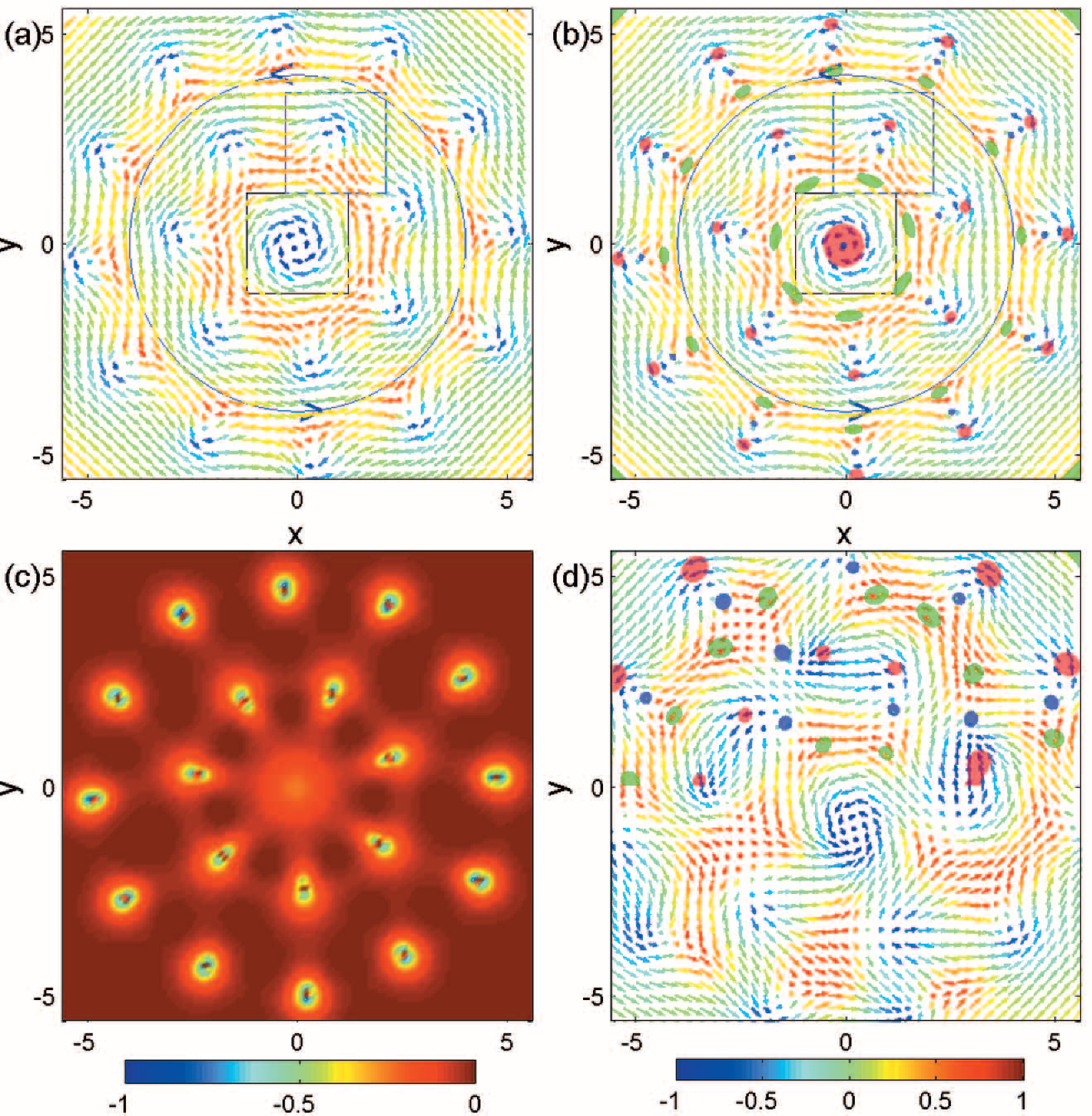}
\caption{\small (Color online). (a) Spin texture of spinor BEC of $^{87}$Rb with $\kappa=0.5$ and $\Omega=0.5\omega$. The color of each arrow indicates the magnitude of $S_{z}$. The black pane points out a Skyrmion,
and the
blue pane indicates a half-Skyrmion. The blue arrows show the main direction of the spin texture. (b) The position of vortices and the spin texture.
The green,
blue and red spots are the center of vortices formed by
the $m_{F}=-1$, $m_{F}=0$ and $m_{F}=+1$ components,
respectively.
(c) Topological charge density.
(d) The position of vortices and the spin texture of spinor BEC of $^{87}$Rb with $\kappa=0.1$ and $\Omega=0.5\omega$.
Noting, we only mark the vortices in $y>0$ region in order to illuminate the spin texture and position of vortices clearly.
The units of length and strength of SOC are $\sqrt{\frac{\hbar}{m\omega}}$, $\sqrt{\hbar\omega/m}$, respectively.
}
\end{figure*}
\par

\section{\bf Results and explanations}

\subsection{A half-Skyrmion in spinor BEC with spin-orbit coupling}
We firstly present a way to obtain a single half-Skyrmion (non-meron-antimeron pair) in the rotating spin-1 BEC with SOC. We begin with
the spinor BEC of $^{87}$Rb \cite{ zhoufei}, which is ferromagnetic (FM) ($g_{s}<0$).
We set our model with $\mu_{j,0}(j=0,\pm1)=1.1\hbar\omega$, $\mu=8\hbar\omega$, $\kappa_x=0.2$, $\kappa_y=1$ and the rotating frequency $\Omega=0.5\omega$. Figure 1(a) displays the densities and phases obtained under the equilibrium state. Obviously, there are several vortices in each component.

The spin texture \cite{ Mizushima, shima, Kasamatsu, matsu} is defined by
\begin{eqnarray} \label{sss}
\textbf{S}_{\alpha}=\sum_{m,n=0,\pm1}\Psi^{*}_{m}(\widehat{F}_{\alpha})_{m,n}\Psi_{n}/|\mathbf{\Psi}|^{2} (\alpha=x,y,z).
\end{eqnarray}
Figure 1(b) shows the corresponding spin texture according to Eqs. (3).
The direction of arrow changes suddenly and the arrows
form a small half circle located near the position $(-2, 0)$. Additionally, the orientations of the arrows suffer a $180^{\circ}$
reversal along the $x$ axis.
For clarification, Figure 1(c) plots an enlarged picture of the special structure.
Note that, for illuminating the relationship between the structure and the vortices,
we have added the position of the vortex in the three components with the color spots.
We find the structure can be represented by the form
\begin{eqnarray} \label{sss}
\textbf{s}_{0x}&=&-\frac{y}{\sqrt{x^{2}+y^{2}}}\exp[-\lambda(x^{2}+y^{2})],\notag \\
\textbf{s}_{0z}&=&\frac{x}{\sqrt{x^{2}+y^{2}}}\exp[-\lambda(x^{2}+y^{2})], \\
\textbf{s}_{0y}&=&-\sqrt{1-\textbf{s}_{0x}^{2}-\textbf{s}_{0z}^{2}}, \notag
\end{eqnarray}
where $\lambda$ is a variable parameter. Clearly, the two vectors $\textbf{s}_{0x}$, $\textbf{s}_{0z}$ can vary from $-1$ to $1$, but the vector $\textbf{s}_{0y}$ varies from $-1$ to $0$. This means that the spin vector can cover half of a unit sphere.
By calculating the topological charge $Q=\frac{1}{4\pi}\int\int \textbf{s}\cdot(\frac{\partial \textbf{s}}{\partial x}\times\frac{\partial \textbf{s}}{\partial y})dxdy$, where $\textbf{s}=\textbf{S}/|\textbf{S}|$ and $\textbf{S}$ comes from Eqs. (3), we find the topological charge approaches $|Q|=0.5$.
Noting the value of the topological charge with Eqs. (4) is $|Q|=0.5$.
If we perform a transformation:
$(\textbf{S}^{'}_{x}, \textbf{S}^{'}_{y}, \textbf{S}^{'}_{z})=(\textbf{S}_{x}, \textbf{S}_{z}, \textbf{S}_{y})$,
we can obtain a clear Skyrmion-like structure [see Fig. 1 (d)].
Meanwhile, we can prove that the transformation does not affect the value of the topological charge $|Q|$ at all.
The explicit proofs are given in Appendix.
Thus, the spin texture in Fig. 1(c) is a half-Skyrmion.

In Refs. \cite{ Kasamatsu, matsu}, Kasamatsu \emph{et al.} have studied the meron-antimeron pair in the two-component BECs without SOC. They have provided expressions to characterize the meron-antimeron pair.
In Ref. \cite{sushiwei}, Su \emph{et al.} have also obtained the meron-antimeron pair by simulating the non-rotating spinor BEC of $^{87}$Rb with SOC. Our study indicates that in the rotating BECs with SOC the half-Skyrmion (meron) can appear without pairing.
Meanwhile, the solution [Eqs. (4)] is completely different from that in Ref. \cite{ Kasamatsu}.
We remark that the above transformation exchanges the spin vectors in Fig. 1(d). It
mainly changes the perspective but not affects the essence of the spin texture.

Figure 1(e) indicates the local density distribution in the region of the half-Skyrmion.
The green, blue and red surfaces represent the densities of
the $m_{F}=-1$, $m_{F}=0$ and $m_{F}=+1$ components, respectively. The cyan
represents the total density.
The $m_{F}=+1$ component forms
an obvious hump in the vortex region of $m_{F}=-1$ component, and vice versa for the $m_{F}=-1$ component.
These properties cause a dipole of spin.
There is a local density minimum at the position of the
vortex formed by the $m_{F}=0$ component when we examine the total density.
This point is different from the normal coreless vortex \cite{ Mizushima, shima, Kasamatsu, matsu, Mermin, Anderson} where the total density has no singularity.
In fact, this structure can be viewed as a three-vortex structure. In the following text, we will show that the three-vortex structure plays an important role in creating the half-Skyrmion.

Figure 2 plots a half-Skyrmion with Eqs. (4). Comparing it with Fig. 1(c), the formation of the two spin texture is very similar.
This indicates the half-Skyrmion can be well described by Eqs. (4) in our simulation.
We obtain Eqs. (4) to characterize the half-Skyrmion intuitively according to the properties of Fig. 1(c).
All the arrows in Fig. 1(c) tend to point down, which means the spin vector $\textbf{s}_{0y} \leq 0$. Especially the arrows completely point down when the position is far away from $(-2,0)$, this properties mean that $\textbf{s}_{0x}$ and $\textbf{s}_{0z}$ should approach 0 when the position is far away from the half-Skyrmion.
Thus, there must be an index function in the expression of the spin-vectors $\textbf{s}_{0x}$ and $\textbf{s}_{0z}$, respectively.
Furthermore, we have found that the half-Skyrmion is related to a three-vortex structure, which causes a dipole of spin. This feature implies that $\textbf{s}_{0x}$ and $\textbf{s}_{0z}$ would be antisymmetrical. With lots of tests on the topological charge and others, we find Eqs. (4) can characterize the half-Skyrmion.
Noting there is a displacement of position between the two plots [Fig. 1(c) and Fig. 2].

\par

\subsection{Half-Skyrmion lattice in rotating spin-1 BEC with spin-orbit coupling}
To obtain the half-Skyrmion lattice, we choose the parameters: $\mu_{j,0}(j=0,\pm1)=3.6\hbar\omega$, $\mu=25\hbar\omega$, $\kappa_x=\kappa_y=\kappa$ and the rotating frequency $\Omega=0.5\omega$. Figure 3 displays the densities and phases obtained under various strengths of SOC.
For a very weak SOC ($\kappa=0.1$), the patterns are irregular. When $\kappa$ is over 0.1, the patterns are relatively regular.
We take the $m_{F}=1$ component as an example. In Fig. 3(b), there are 8 the nearest vortices around the center vortex of the $m_{F}=1$ component.
The number is 7, 3 and 4 in Fig. 3(c), 3(d) and 3(f) respectively.
Thus, these pictures factually display the transition of the patterns as the strength of SOC increases.
SOC can be used to adjust the pattern of the rotating spin-1 BEC.

Just as previous experiments about
the rotating BECs \cite{Bradley, Su}, there are some
vortices in the three components respectively.
The fifth and sixth columns indicate the phases of $m_{F}=-1$ and $m_{F}=1$ components respectively.
Like the vortex lattice in the single-component BEC, there are some lines where the phases change discontinuously from red to blue, which corresponds to the branch cuts between the phases $-\pi$ and $\pi$. The ends represent phase defects.
All the lines extend to the outskirts of the BEC where the density of the BEC is almost negligible, and end with another defect which offers neither the energy nor the angular momentum to the system.
Furthermore, we also
find some peaks accompanying the vortices, regularly arraying to be triangle, square, heptagon \emph{etc}, especially in
the center of the $m_{F}=-1$ component.

Unlike the periodic vortex lattice \cite{ Bradley} or the vortices trimers \cite{ Su}, the vortices encircle the center with several circles.
The number of vortices is 1 or 0 in the center, and it increases as the radius increases. Certainly,
this phenomenon is not obvious when SOC is very weak ($\kappa=0.1$).
The fourth column shows
the total density of BECs. Here, we can distinguish some local
minimum of densities, especially when $\kappa$ approaches 1.

Two components of the BECs have a corresponding vortex at the center and the other component forms a hump filled in the center vortices.
In addition, one of the center vortex is formed by the $m_{F}=1$ component.
For example, the $m_{F}=0$ and $m_{F}=1$ components have a corresponding center vortex in Fig. 3(b), 3(c) and 3(e). But, in Fig. 3(d), the center vortex is created in the $m_{F}=-1$ and $m_{F}=1$ components, respectively. The center vortices is formed in which two components depends on lots of factors such as the number of vortex, number of atoms of each component, rotation frequency and the strength of SOC.
Generally, if the center vortex of the $m_{F}=1$ component is a multiply charged vortex, the other center vortex will occur in the $m_{F}=0$ component.
Otherwise, the other center vortex will occur in the $m_{F}=-1$ component.

\begin{figure}[thbp]
\includegraphics[width=9cm,trim=0in 0.0in 0in 0.0in]{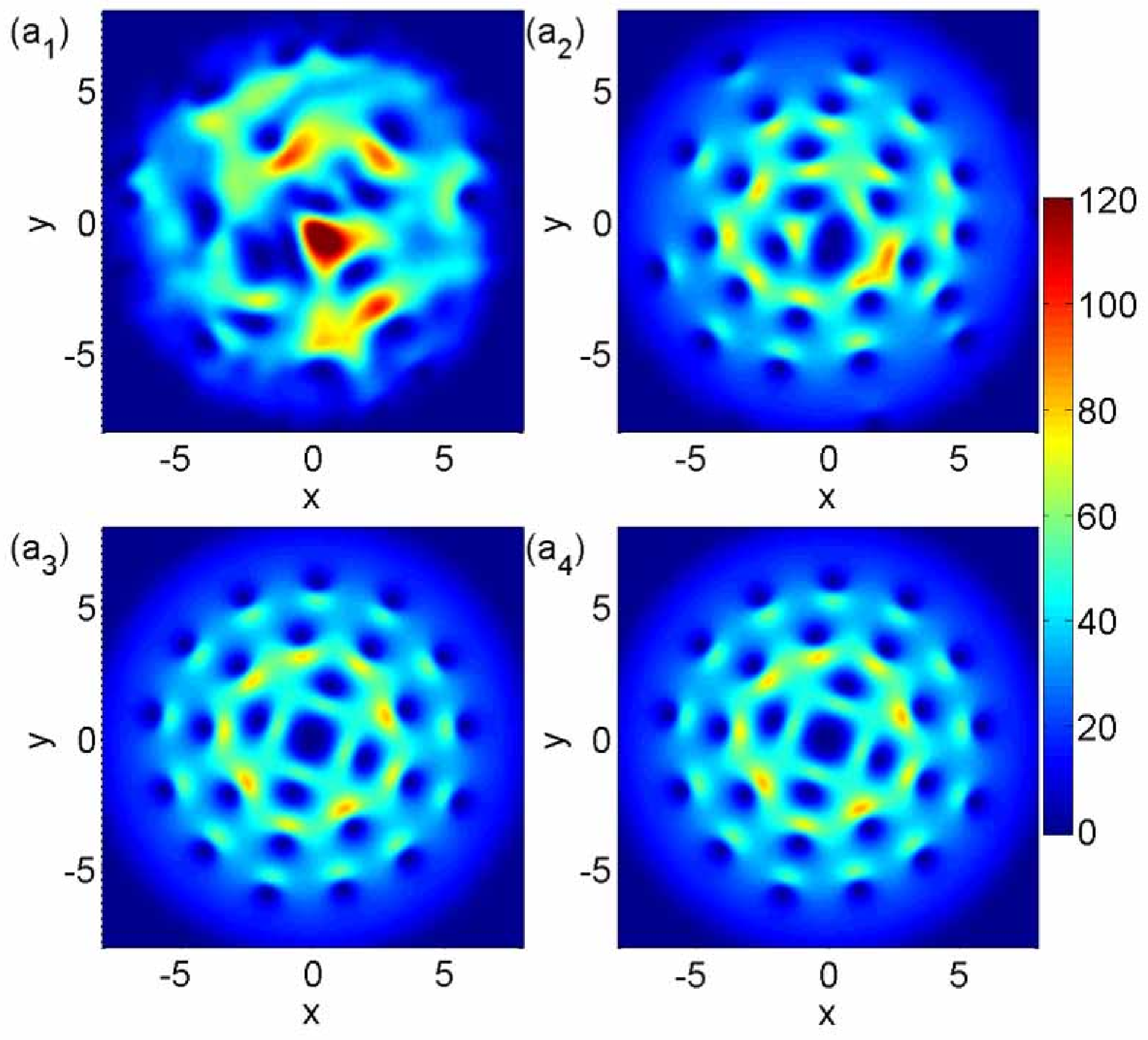}
\includegraphics[width=9cm,trim=0in 0.1in 0in 0.2in]{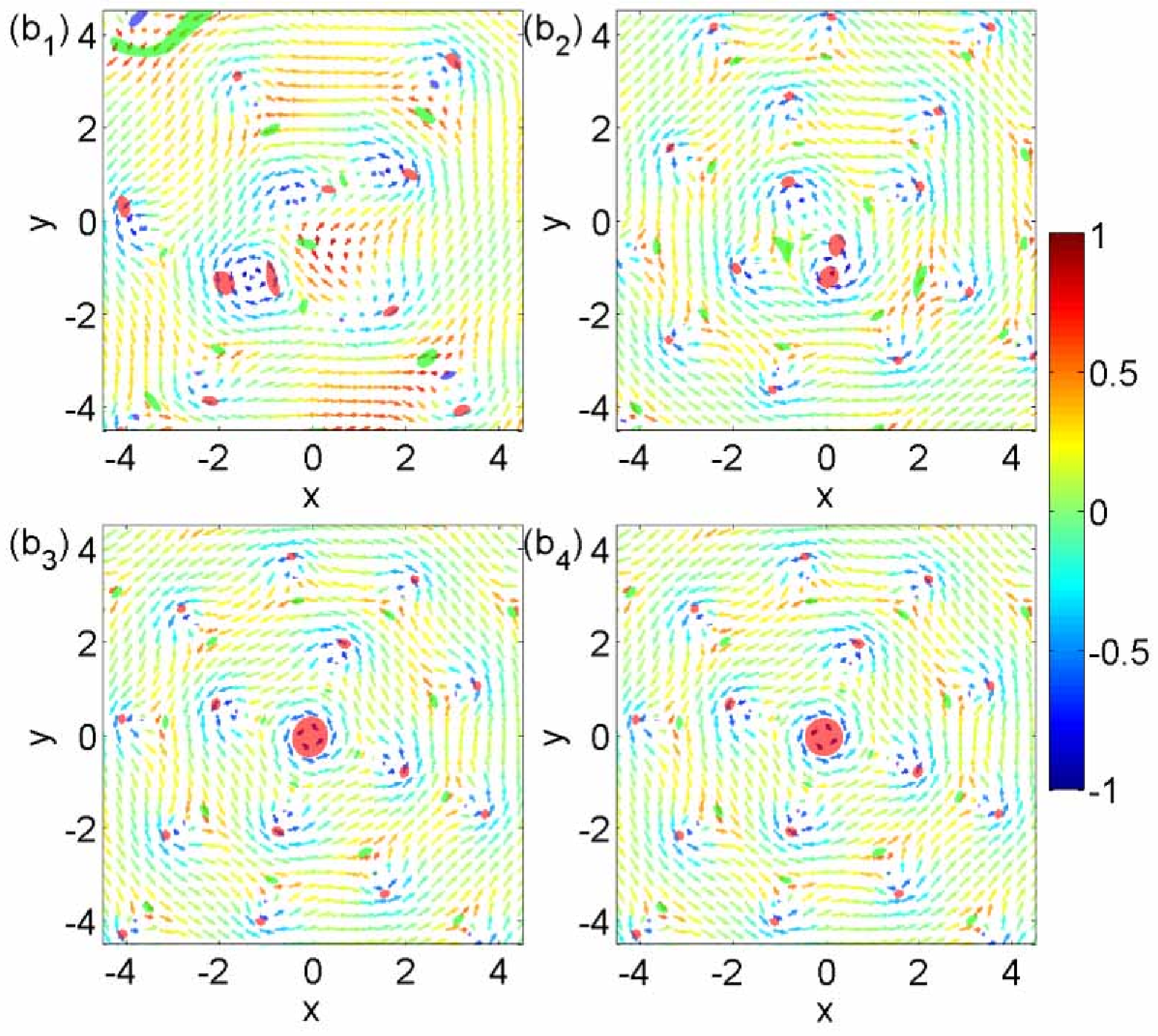}
\caption{\small (Color online). The time evolution of the spin-1 BEC of $^{87}$Rb with $\mu_{j,0}(j=0,\pm1)=3.6\hbar\omega$, $\mu=25\hbar\omega$, $\kappa_x=\kappa_y=1$ and $\Omega=0.5\omega$.
(a$_{1}$)-(a$_{4}$) show the densities of the $m_{F}=1$ component at $t=20\omega^{-1}$, $60\omega^{-1}$, $200\omega^{-1}$ and $500\omega^{-1}$, respectively; (b$_{1}$)-(b$_{4}$) show the corresponding spin texture and the position of vortices. The meanings of the spots and the colored arrows are the same as those in Fig. 4(b). The units of length and strength of SOC are $\sqrt{\frac{\hbar}{m\omega}}$, $\sqrt{\hbar\omega/m}$, respectively.
}
\end{figure}

\begin{figure*}[tbp]
\includegraphics[width=14.5cm]{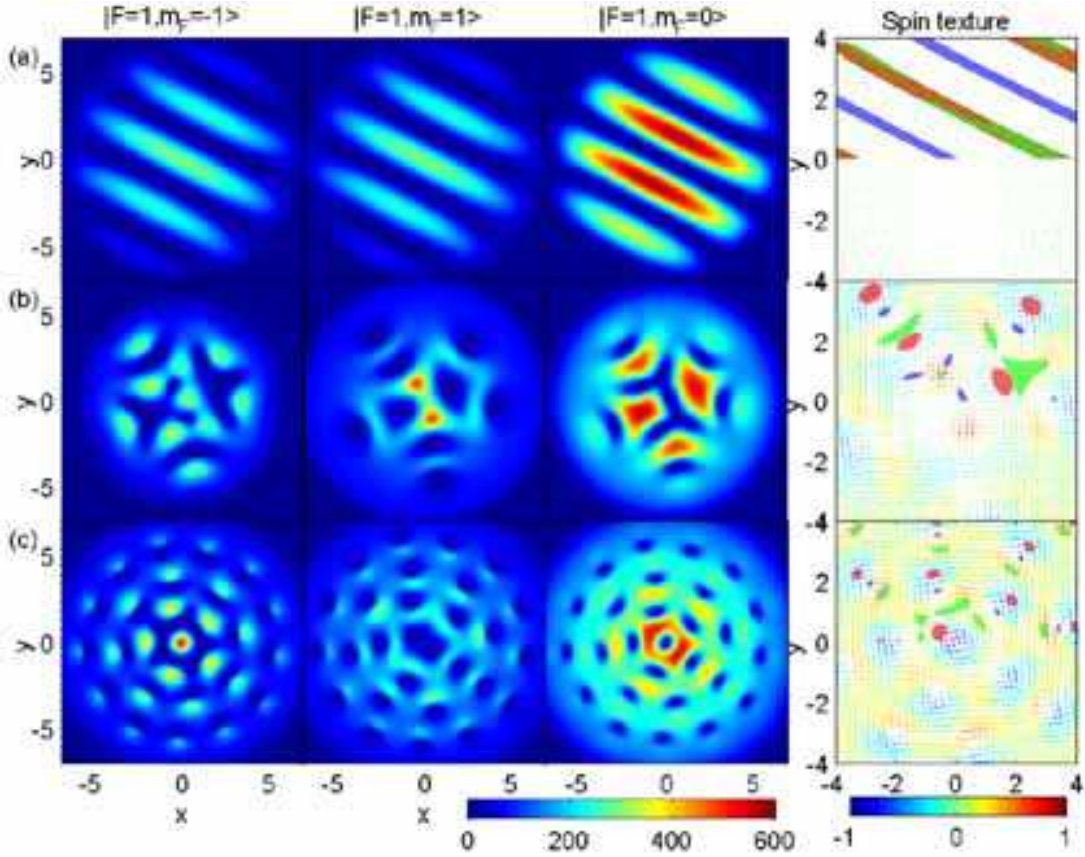}
\caption{\small (Color online). The effect of rotation frequency $\Omega$ for spinor BEC of $^{23}$Na with $\mu_{j,0}(j=0,\pm1)=3.6\hbar\omega$, $\mu=25\hbar\omega$, $\kappa_x=\kappa_y=\kappa=1$, $a_{0}=50a_{B}$ and $a_{2}=55a_{B}$. (a) $\Omega=0$; (b) $\Omega=0.2\omega$; (c) $\Omega=0.5\omega$. The fourth column shows the corresponding spin textures and position of vortices. The meanings of the spots and the colored arrows are the same as those in Fig. 4(b).
Noting, we only mark the vortices in $y>0$ region. The atom numbers ($N_{-1}$, $N_{1}$, $N_{0}$) approximately are ($1.2\times10^{4}$, $1.2\times10^{4}$, $2.36\times10^{4}$), ($1.13\times10^{4}$, $1.5\times10^{4}$, $2.46\times10^{4}$), and ($1.67\times10^{4}$, $2.0\times10^{4}$, $3.53\times10^{4}$), respectively. The units of length and strength of SOC are $\sqrt{\frac{\hbar}{m\omega}}$, $\sqrt{\hbar\omega/m}$, respectively.}
\end{figure*}
\par

\begin{figure}[tbp]
\includegraphics[width=8.5cm,trim=0in 0.2in 0in 0.3in]{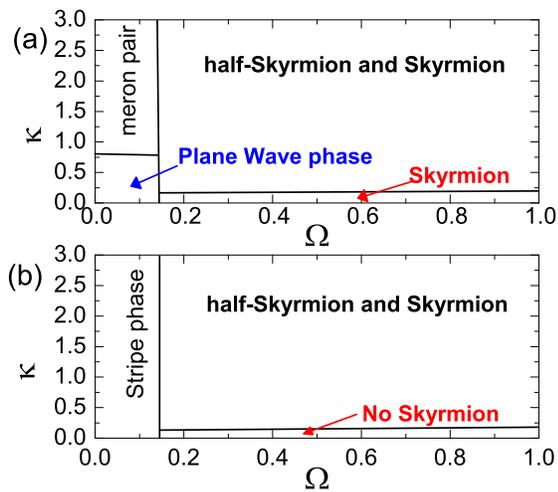}
\caption{\small (Color online). The phase diagrams of the products in spin-1 BEC with the rotation frequency $\Omega$ and the SOC strength $\kappa$ ($\kappa_x=\kappa_y=\kappa$). Here, $\mu_{j,0}(j=0,\pm1)=3.6\hbar\omega$, $\mu=25\hbar\omega$. (a) $^{87}$Rb; (b) $^{23}$Na. The units of rotation frequency and strength of SOC are $\omega$, $\sqrt{\hbar\omega/m}$, respectively.}
\end{figure}
\par

\begin{figure*}[thbp]
\includegraphics[width=8.5cm]{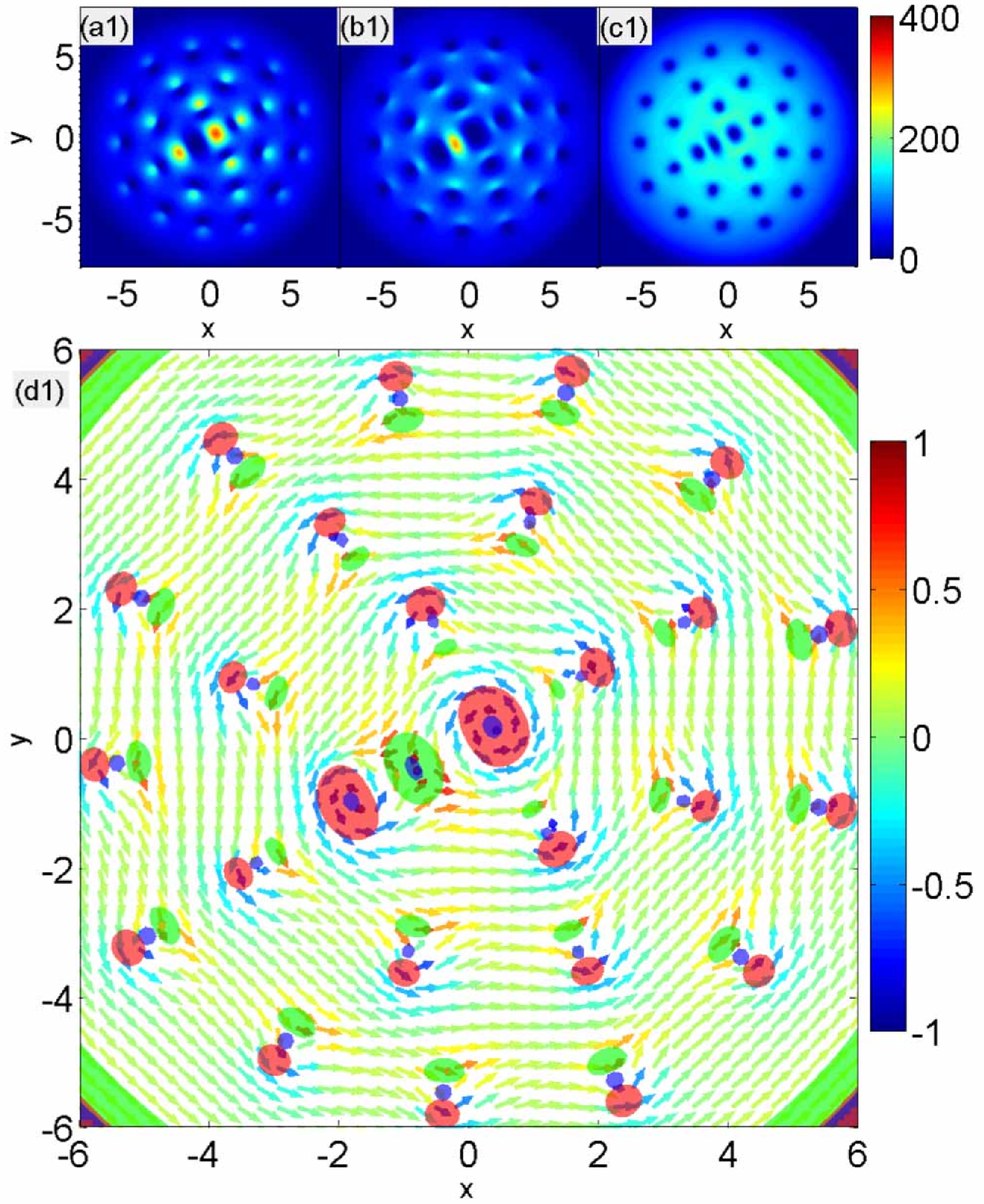}
\includegraphics[width=8.5cm]{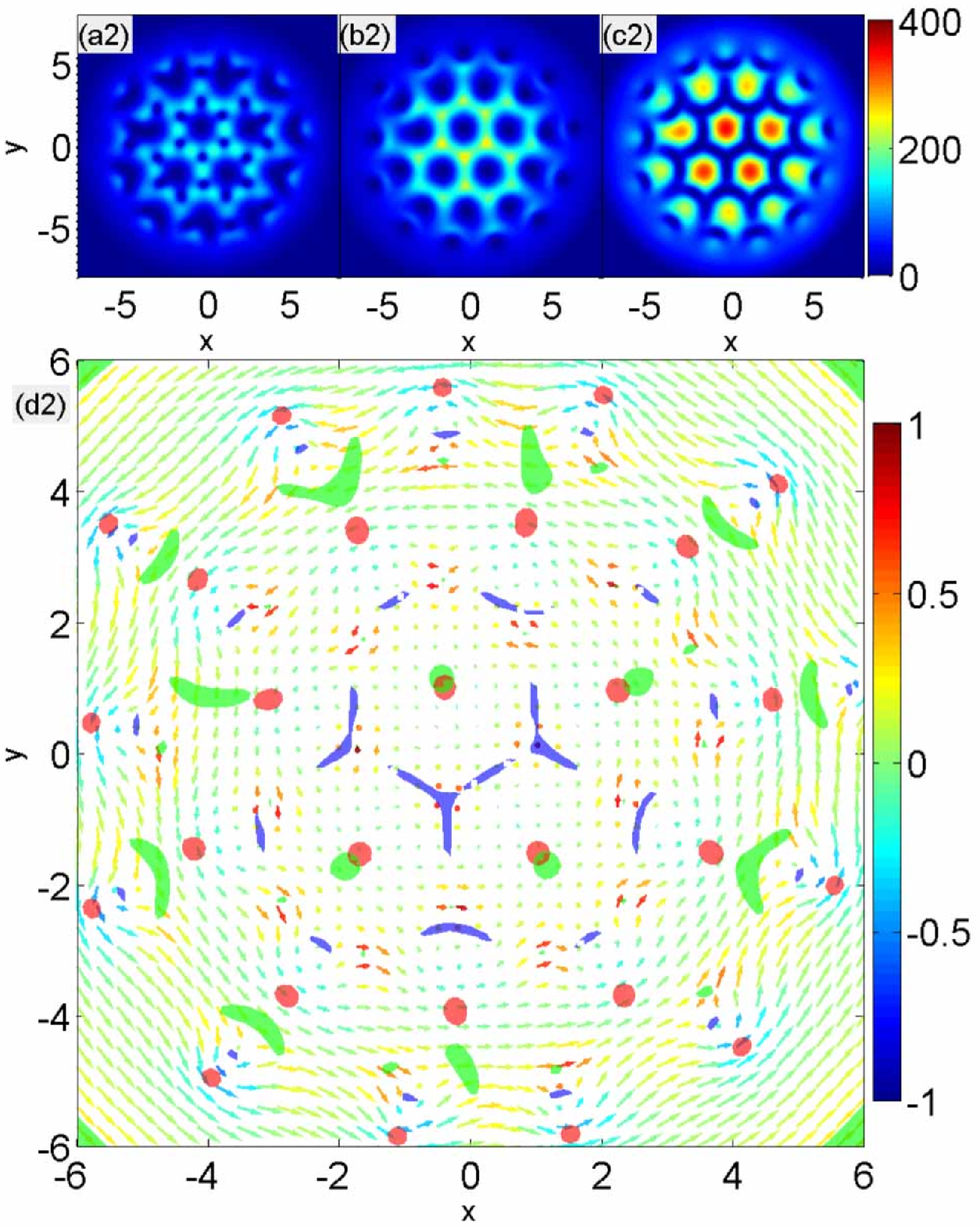}
\caption{\small (Color online). The effect of tuning ferromagnetic and antiferromagnetic interactions for BEC with $\mu_{j,0}(j=0,\pm1)=3.6\hbar\omega$, $\mu=25\hbar\omega$, $\kappa_x=\kappa_y=1$ and $\Omega=0.5\omega$. (a1), (b1) and (c1) show the densities of the $m_{F}=-1$, $m_{F}=1$ and $m_{F}=0$ components of $^{87}$Rb with $a_{0}=101.8a_{B}$ and $a_{2}=50.2a_{B}$ respectively ($N_{-1}\approx1.0\times10^{4}$, $N_{1}\approx1.1\times10^{4}$, $N_{0}\approx1.91\times10^{4}$) (strong ferromagnetic case); (d1) shows the spin texture and the position of vortices. The meanings of the spots and the colored arrows are the same as those in Fig. 4(b). (a2)-(d2) indicate the corresponding results of $^{23}$Na with $a_{0}=50a_{B}$ and $a_{2}=110a_{B}$
($N_{-1}\approx9.2\times10^{3}$, $N_{1}\approx1.2\times10^{4}$, $N_{0}\approx1.75\times10^{4}$) (strong antiferromagnetic case).
The units of length and strength of SOC are $\sqrt{\frac{\hbar}{m\omega}}$, $\sqrt{\hbar\omega/m}$, respectively.
}
\end{figure*}

There are many half-Skyrmion rounding the center several
circles, i.e., radially arranging in the system [see Fig. 4(a)].
Meanwhile,
the arrows form big rings, whose main direction is marked with the blue arrows.
Figure 4(b) illuminates the relationship
between the vortices and the half-Skyrmions.
Except for the central area of the BECs, we find the
positions of vortices in the three components are far away from the
center with the order: green, blue and red. Meanwhile, the sequence distributes in the whole system.
Furthermore, Figure 4(c) plots the topological charge density, which is defined by
$q(r)=\frac{1}{4\pi} \textbf{s}\cdot(\frac{\partial \textbf{s}}{\partial x}\times\frac{\partial \textbf{s}}{\partial y})$, where $\textbf{s}=\textbf{S}/|\textbf{S}|$.
The value of the topological charge density varies only from $-1$ to $0$. This property indicates that our results in Fig. 4(a) are half-Skyrmions but not meron-antimeron pairs.
Obviously, each half-Skyrmion with the nontrivial topological charge density accompanies a three-vortex structure. Thus, we can view the three-vortex structure as a cell.
Undoubtedly, the number of
vortices in three components approaches $1:1:1$.
Usually, the Skyrmion-like excitations \cite{ Mizushima, shima, Kasamatsu, matsu} are related to the underlying vortex configuration such as Mermin-Ho \cite{ Mermin} and Anderson-Toulouse \cite{ Anderson} vortices.
Here, the half-Skyrmion is related to the three-vortex structure.
The distribution of half-Skyrmion depends on that of vortices.
Thus, it is not inconceivable that the different strength of SOC will
cause various half-Skyrmion lattices.

In the absence of SOC, the periodic Skyrmion lattice can be created in the rotating spin-1 BEC of $^{87}$Rb \cite{ Su}.
Figure 4(d) shows the spin textures and the position of vortices
when the strength of SOC is 0.1.
The vortices hardly form the three-vortex structure, especially near the center.
Additionally, the half-Skyrmion lattice is not very obvious.
These results further prove that the half-Skyrmion is related to the three-vortex structure.
To obtain the half-Skyrmion lattice, the strength of SOC must exceed a critical value. Here, this value approaches 0.2.
Noting, we do not fix the ratio of the three components in the dynamical process.
The mixture ratio of the three components depends on the system itself.

Figure 5 indicates the time evolution of the system. Here, we take $^{87}$Rb with $\mu_{j,0}(j=0,\pm1)=3.6\hbar\omega$, $\mu=25\hbar\omega$, $\kappa_x=\kappa_y=1$ and $\Omega=0.5\omega$ as an example. Fig. 5(a$_{1}$)-5(a$_{4}$) only indicate the densities of $m_{F}=1$ component at
$t=20\omega^{-1}$, $60\omega^{-1}$, $200\omega^{-1}$ and $500\omega^{-1}$, respectively. The density distribution
in Fig. 5(a$_{1}$), Fig. 5(a$_{2}$) and Fig. 5(a$_{3}$) are different from each other. However, the density distribution in Fig. 5(a$_{3}$) is almost the same as that in Fig. 5(a$_{4}$).
These properties mean that the system has reached the equilibrium state and the density distribution is dynamically stable.
Similarly, the $m_{F}=-1$ and $m_{F}=0$ components also have these properties.
Fig. 5(b$_{1}$)-5(b$_{4}$) are the corresponding spin texture at $t=20\omega^{-1}$, $60\omega^{-1}$, $200\omega^{-1}$ and $500\omega^{-1}$, respectively.
Obviously, the spin texture is stable when the system reaches the equilibrium state.
In fact, all the results are dynamically stable when the system reaches the equilibrium state under the quenching process.

\subsection{The effect of the rotation frequency}
We find the half-Skyrmion lattice can also occur in the antiferromagnetic (AFM) BEC, where $g_{s}>0$ [see Fig. 6(c)].
There is no Skyrmion excitation appearing in rotating AFM BEC of $^{23}$Na \cite{ Su} when $\kappa=0$.
Here, we use the spinor BEC of $^{23}$Na to illustrate the effect of the rotation frequency. We only change
the rotation frequency $\Omega$ and fix all other parameters to perform the numerical experiments.
Figure 6 shows the density distribution and spin texture under various rotation frequencies.
In the absence of rotation ($\Omega=0$), there is no vortex appearing at all. Each component of the BECs is split into several parallel parts.
In fact, these properties agree with the stripe phase \cite{Wang, TinLun, Jian, Sinha}. The spin texture indicates no half-Skyrmion excitations in this system.
Color straps factually are the low density domain of BECs.
Added a weak rotation ($0.2\omega$), the splitting parts bend and break, and several vortices and the three-vortex structure occur. When the rotation becomes faster ($0.5\omega$), the vortex lattice emerges and the half-Skyrmion lattice is very obvious. Naturally,
the rotation can control the half-Skyrmion lattice because it can induce the underlying vortices.

Now, we can systematically understand the half-Skyrmion phenomenon in the rotating spinor BECs.
Its dynamics is driven by
the rotation in the quenching process and the intrinsic
spin-Hall effect derived from the effective SOC.
As is well
known in the study of rotating superfluid helium \cite{ Campbell, Tsubota}, the
rotating drive pulls vortices into the rotation axis, while repulsive interaction tends to push them apart; this competition
yields a vortex lattice whose vortex density depends on the
rotation frequency. Meanwhile, SOC causes the spin separation and creates the dipole structure of spin, which is embedded in the three-vortex structure.
The half-Skyrmion derives from the dipole structure. In term of densities, only a core
structure appears in the center. That is the origin of the center
Skyrmion.

Figure 7 plots the phase diagrams of the spinor BECs with various products in our experiments. In this paper, we mainly focus on the half-Skyrmion as well as the center Skyrmion. The half-Skyrmion phase, where the center Skyrmion may appear, occupies a large region of the $\Omega-\kappa$ plane in both the FM BEC and the AFM BEC. For FM BEC of $^{87}$Rb, the meron-antimeron pairs \cite{sushiwei} can occur when SOC is greater than a critical value ($\approx0.8$) and the rotation is very weak. Figure 7(a) also shows the condition for obtaining the plane wave phase ($\kappa \leq 0.8$, $\Omega \leq 0.15 \omega$) and for creating the Skyrmion lattice ($\kappa \leq 0.2$, $\Omega > 0.15 \omega$).
For AFM BEC of $^{23}$Na, the Skyrmion or half-Skyrmion hardly occurs when $\kappa$ is smaller than 0.2 [see Fig. 7(b)].
If the rotation is very weak ($\Omega \leq 0.15 \omega$), the stripe phase will be observed.

\subsection{The effect of tuning the ferromagnetic and antiferromagnetic interactions}
Generally speaking, $g_{s}$ is much smaller than $g_{n}$.
By adjusting the two $s$-wave scattering lengths $a_{0}$ and $a_{2}$ through Feshbach resonances, the spin exchange interaction strength $g_{s}$ is tunable.
We now perform the above experiments by changing $a_{2}$. Figures 8(a1)-(d1) show a stronger FM case of $^{87}$Rb ($g_{s}/g_{n}=-0.255$). Contrasting the densities and the spin texture, we find the three-vortex structure and the half-Skyrmion lattice are common in this FM BEC, though there are several half-Skyrmions of the circular/hyperbolic structure near the center. Furthermore, we also test a stronger AFM BEC of $^{23}$Na ($g_{s}/g_{n}=0.222$). The three-vortex structure and the half-Skyrmion mainly emerge at the outskirt of the BECs. In fact, the strong AFM interactions restrict the half-Skyrmion excitations, so the half-Skyrmion hardly emerges in the center. If we increase the strength of SOC, we will obtain a more obvious picture about half-Skyrmion lattice. These experiments show
the universality of the half-Skyrmion excitation in BEC with SOC.

\par

\section{\bf Conclusion}
We have studied the half-Skyrmion in the rotating spin-1 BEC with SOC.
We find the half-Skyrmions (meron) can occur but not forms the meron-antimeron pairs.
This phenomenon implies a new understanding of the half-Skyrmion in BECs.
The half-Skyrmion originates from a dipole resulted from a local spin separation. Meanwhile, the half-Skyrmion excitation is related to a three-vortex structure where the dipole is embedded. The half-Skyrmion excitation can occur as long as the three-vortex structure appears, even in the strong FM and AFM BEC with SOC.
The provided phase diagrams indicate the condition of obtaining the half-Skyrmions in spinor BEC of both $^{87}$Rb and $^{23}$Na.
Our study gives an experimental protocol to observe these novel phenomena in future experiments. Not only do our findings exist in spin-1 BEC, but also the related textures should appear in high-spin BEC, superfluid and superconduction. This work is of particular significance for exploring the novel topological excitation such as half-Skyrmions in quantum gas and condensed matter physics.

\par
\section*{Acknowledgments}

We are grateful to S.-C. Gou for useful comments.
This work was supported by the NKBRSFC under grants Nos. 2011CB921502,
2012CB821305, 2009CB930701, 2010CB922904, and NSFC under grants
Nos. 10934010, 60978019, 11001263 and NSFC-RGC under grants Nos. 11061160490 and 1386-N-HKU748/10.

\section*{Appendix: The numerical proofs on the unchanged topological charge $|Q|$ under the transformation $(\textbf{S}^{'}_{x}, \textbf{S}^{'}_{y}, \textbf{S}^{'}_{z})=(\textbf{S}_{x}, \textbf{S}_{z}, \textbf{S}_{y})$}

We now prove that the topological charge $|Q|$ is unchanged under the transformation $(\textbf{S}^{'}_{x}, \textbf{S}^{'}_{y}, \textbf{S}^{'}_{z})=(\textbf{S}_{x}, \textbf{S}_{z}, \textbf{S}_{y})$. It is well known that the topological charge $Q$ is defined as
\begin{eqnarray} \label{sss}
Q=\frac{1}{4\pi}\int\int \textbf{s}\cdot(\frac{\partial \textbf{s}}{\partial x}\times\frac{\partial \textbf{s}}{\partial y})dxdy,
\end{eqnarray} \label{sss}
where $\textbf{s}=\textbf{S}/|\textbf{S}|$. For clarity, we use $Q(\textbf{s}_{x}, \textbf{s}_{y}, \textbf{s}_{z})$ to describe the normal expression of the topological charge $Q$ of Eq. (5), and $Q(\textbf{s}_{x}, \textbf{s}_{z}, \textbf{s}_{y})$ to denote the topological charge under the transformation $(\textbf{S}^{'}_{x}, \textbf{S}^{'}_{y}, \textbf{S}^{'}_{z})=(\textbf{S}_{x}, \textbf{S}_{z}, \textbf{S}_{y})$.

We can expand the expression $Q(\textbf{s}_{x}, \textbf{s}_{y}, \textbf{s}_{z})$ to be the form
\begin{eqnarray}
Q(\textbf{s}_{x}, \textbf{s}_{y}, \textbf{s}_{z})=\frac{1}{4\pi} \int\int \left|
  \begin{array}{ccc}
    \textbf{s}_{x} & \textbf{s}_{y} & \textbf{s}_{z} \\
    \frac{\partial}{\partial x} \textbf{s}_{x} & \frac{\partial}{\partial x} \textbf{s}_{y} & \frac{\partial}{\partial x} \textbf{s}_{z} \\
    \frac{\partial}{\partial y} \textbf{s}_{x} & \frac{\partial}{\partial y} \textbf{s}_{y} & \frac{\partial}{\partial y} \textbf{s}_{z} \\
  \end{array}
\right| dx dy. \notag
\end{eqnarray}

Similarly, $Q(\textbf{s}_{x}, \textbf{s}_{z}, \textbf{s}_{y})$ can be written as
\begin{eqnarray}
Q(\textbf{s}_{x}, \textbf{s}_{z}, \textbf{s}_{y})=\frac{1}{4\pi} \int\int \left|
  \begin{array}{ccc}
    \textbf{s}_{x} & \textbf{s}_{z} & \textbf{s}_{y} \\
    \frac{\partial}{\partial x} \textbf{s}_{x} & \frac{\partial}{\partial x} \textbf{s}_{z} & \frac{\partial}{\partial x} \textbf{s}_{y} \\
    \frac{\partial}{\partial y} \textbf{s}_{x} & \frac{\partial}{\partial y} \textbf{s}_{z} & \frac{\partial}{\partial y} \textbf{s}_{y} \\
  \end{array}
\right| dx dy \notag
\end{eqnarray}
\begin{eqnarray}
    =-\frac{1}{4\pi} \int\int \left|
  \begin{array}{ccc}
    \textbf{s}_{x} & \textbf{s}_{y} & \textbf{s}_{z} \\
    \frac{\partial}{\partial x} \textbf{s}_{x} & \frac{\partial}{\partial x} \textbf{s}_{y} & \frac{\partial}{\partial x} \textbf{s}_{z} \\
    \frac{\partial}{\partial y} \textbf{s}_{x} & \frac{\partial}{\partial y} \textbf{s}_{y} & \frac{\partial}{\partial y} \textbf{s}_{z} \\
  \end{array}
\right| dx dy=-Q(\textbf{s}_{x}, \textbf{s}_{y}, \textbf{s}_{z}). \notag
\end{eqnarray}

Thus, we obtain $Q(\textbf{s}_{x}, \textbf{s}_{y}, \textbf{s}_{z})=-Q(\textbf{s}_{x}, \textbf{s}_{z}, \textbf{s}_{y})$, and $|Q(\textbf{s}_{x}, \textbf{s}_{y}, \textbf{s}_{z})|=|Q(\textbf{s}_{x}, \textbf{s}_{z}, \textbf{s}_{y})|$. The topological charge $|Q|$ is unchanged under the transformation $(\textbf{S}^{'}_{x}, \textbf{S}^{'}_{y}, \textbf{S}^{'}_{z})=(\textbf{S}_{x}, \textbf{S}_{z}, \textbf{S}_{y})$.

Similarly, the topological charge $|Q|$ is unchanged, no matter how we exchange the spin vectors $\textbf{S}_{x}$, $\textbf{S}_{y}$ and $\textbf{S}_{z}$.

\end{document}